\documentclass{article}

\usepackage{arxiv}

\usepackage[utf8]{inputenc} 
\usepackage[T1]{fontenc}    
\usepackage{hyperref}       
\usepackage{url}            
\usepackage{booktabs}       
\usepackage{amsfonts}       
\usepackage{nicefrac}       
\usepackage{microtype}      
\usepackage{lipsum}
\usepackage{graphicx}
\graphicspath{ {./images/} }

\usepackage{amsfonts}   
\usepackage{amsmath}    
\usepackage{multirow}   

\title{ViRAC: A Vision-Reasoning Agent Head Movement Control Framework in Arbitrary Virtual Environments}

\author{
 Juyeong Hwang, Seonun Hong \\
  Kyung Hee University\\
  Korea, Republic\\
  \texttt{\{dudyyyy4,zen152\}@khu.ac.kr} \\
   \And
 Hyeongyeop Kang \\
  Korea University\\
  Korea, Republic\\
  \texttt{siamiz\_hkang@korea.ac.kr} \\
}

\begin{document}

\maketitle

\begin{figure}[!h]
    \centering
    \includegraphics[width=\textwidth]{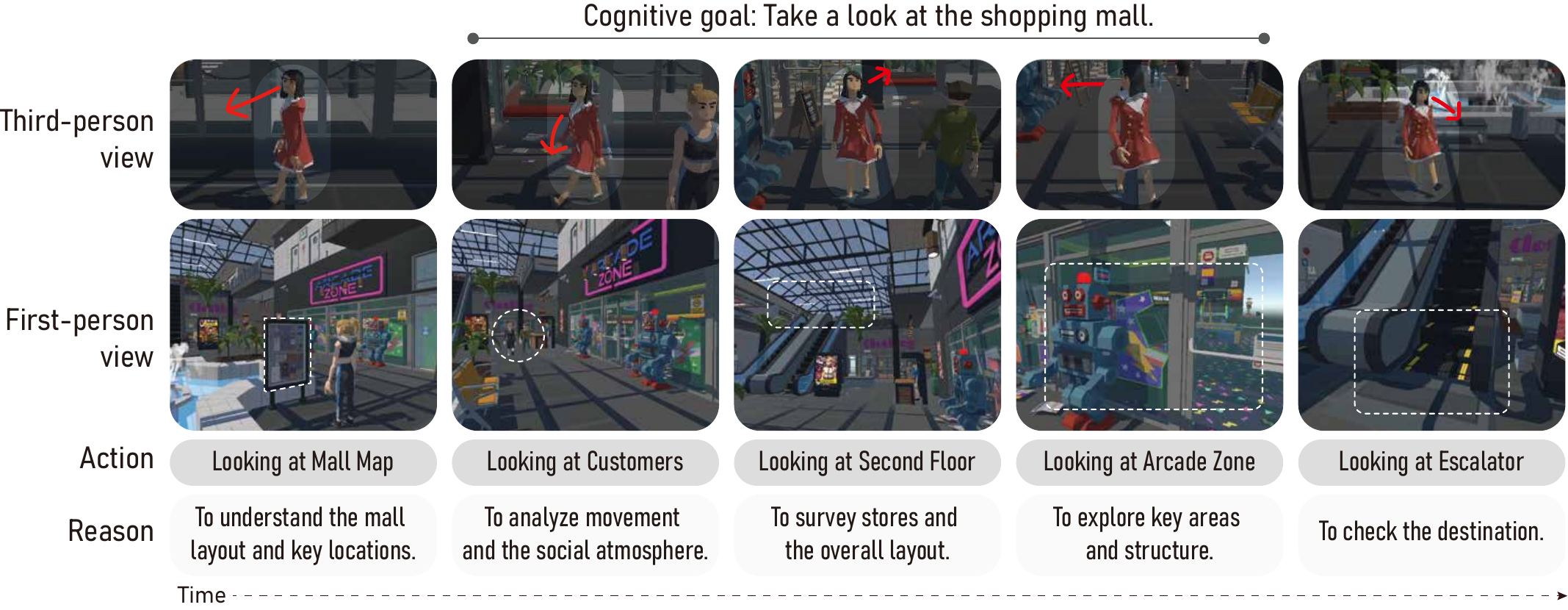} 
    \caption{Visualization example of our framework's head-turn in a busy shopping mall scenario. The top row shows the agent’s third-person view, while the middle row depicts the corresponding first-person view. Each selected action (e.g., ``Looking at Mall Map”) is annotated with the rationale (``To understand the mall layout and key locations”), showing how ViRAC’s cognitive reasoning and visual perception modules interact to produce context-aware head rotations that closely resemble human behavior.}
    \label{fig:teaser}
\end{figure}

\begin{abstract}
Creating lifelike virtual agents capable of interacting with their environments is a longstanding goal in computer graphics. 
This paper addresses the challenge of generating natural head rotations, a critical aspect of believable agent behavior for visual information gathering and dynamic responses to environmental cues. 
Although earlier methods have made significant strides, many rely on data-driven or saliency-based approaches, which often underperform in diverse settings and fail to capture deeper cognitive factors such as risk assessment, information seeking, and contextual prioritization.
Consequently, generated behaviors can appear rigid or overlook critical scene elements, thereby diminishing the sense of realism.
In this paper, we propose \textbf{ViRAC}, a \textbf{Vi}sion-\textbf{R}easoning \textbf{A}gent Head Movement \textbf{C}ontrol framework, which exploits the common-sense knowledge and reasoning capabilities of large-scale models, including Vision-Language Models (VLMs) and Large-Language Models (LLMs). 
Rather than explicitly modeling every cognitive mechanism, ViRAC leverages the biases and patterns internalized by these models from extensive training, thus emulating human-like perceptual processes without hand-tuned heuristics. 
Experimental results in multiple scenarios reveal that ViRAC produces more natural and context-aware head rotations than recent state-of-the-art techniques. Quantitative evaluations show a closer alignment with real human head-movement data, while user studies confirm improved realism and cognitive plausibility. 
\end{abstract}

\section{Introduction}
Realistic and context-aware virtual agents have long been a central research focus in computer graphics, aiming to enhance user immersion in interactive environments such as simulations, games, and virtual reality (VR)~\cite{lerner2010context, steel2010context, narang2016pedvr, curtis2022toward}. One of the key aspects of believable agent behavior lies in natural head rotations—the subtle yet critical movements people make to gather visual information and respond to environmental cues while navigating the real world. By replicating these head movements, virtual agents can provide more lifelike experiences and improve user engagement.

Previous approaches~\cite{xu2018gaze,zhu2020learning, yang2021hierarchical,rondon2022track} to generating head rotations have frequently employed data-driven or saliency map–based techniques. Although these methods capture certain visually prominent elements, their data-driven nature limits adaptability across diverse scenarios, as they often rely on distributions closely aligned with a particular training set~\cite{hsu2020generalized}. Moreover, they fail to incorporate the multifaceted cognitive context that real humans rely on—such as balancing risk assessment, seeking new information, and shifting priorities once information has been obtained. As a result, agents may exhibit awkward behavior, repeatedly focusing on unimportant objects or overlooking potential threats, thereby breaking the sense of realism.

To address these limitations, we propose ViRAC,  a \textbf{Vi}sion-\textbf{R}easoning framework for Realistic \textbf{A}gent Head Rotations that leverages the vision and reasoning capabilities of Vision-Language Models (VLMs) and Large-Language Models (LLMs)~\cite{yang2024qwen2, havrilla2024glore}. Rather than explicitly modeling every possible cognitive factor—such as visual dynamics and human perceptual processes—we capitalize on the rich contextual knowledge that large-scale models acquire from extensive training. Our approach seeks to replicate human-like behaviors in diverse scenarios by utilizing the inherent biases and patterns learned from massive image-text datasets, thus circumventing the need for hand-tuned heuristics.

ViRAC consists of two primary modules: Perception and Decision-making. The Perception Module combines a VLM and a Foundational Memory Module (FMM) to process the agent’s first-person view and maintain an up-to-date record of relevant objects. The VLM autonomously detects and annotates these objects with coherent textual descriptions, while the FMM stores them for extended recall, ensuring seamless continuity even when objects leave and later reenter the agent’s field of view.

The Decision-making Module incorporates an Action History Module (AHM) and an LLM. The AHM logs every action taken by the agent, preserving the semantic structure of behaviors in a human-readable format. By referencing this action history, the LLM decomposes high-level cognitive goals into sub-tasks and selects the next action, balancing exploration with task-focused objectives. 
To further mirror human reasoning, we conducted a user study to collect empirical data on real head-rotation behaviors and their underlying rationales; insights from this study informed the prompts fed to the LLM. 
Through an iterative cycle of perception, reasoning, and environment updates, ViRAC produces dynamic, context-sensitive head rotations.

In summary, our contributions are as follows: 
\begin{itemize}
    \item \textbf{VLM/LLM-Driven Head Rotation}: We are the first to demonstrate how insights learned from large-scale image and text datasets can be harnessed for agent head rotations, without explicitly modeling every nuance of human cognition. 
    \item \textbf{Human Data Collection}: We gather real head-rotation data alongside participants’ stated rationales, providing crucial insight into human cognitive processes for head-movement determination.
    \sloppy
    \item \textbf{Implicit Cognitive Modeling Framework}: We present ViRAC, a framework comprising perception and decision-making modules to simulate the natural human behavior of turning the head for information gathering and dynamic responses to environmental cues.
    \sloppy
    \item \textbf{Broad Applicability and Data-Free Operation}: Our method operates plausibly across a wide range of scenarios without specialized data or task-specific refinements, making it readily adaptable to diverse applications. 
\end{itemize}

\section{RELATED WORK}

Head movement prediction has become a critical component for optimizing user experience in 360-degree video consumption, particularly in head-mounted display (HMD) systems. 
Early methods largely relied on simple trajectory extrapolation—using linear regression or weighted averages—to forecast head orientations~\cite{qian2016optimizing, duanmu2017prioritized}.
Although these techniques proved computationally efficient, they often failed to capture the intricate links between user attention and dynamic video content. 
More recent research highlights the value of integrating user attention signals, typically derived from saliency detection, into prediction models~\cite{rondon2022track}.
This integration not only improves prediction accuracy but also enhances rendering responsiveness.

\cite{nguyen2018your} introduced the concept of panoramic saliency tailored specifically for 360-degree videos. Unlike traditional saliency models that suffer from central bias and multi-object confusion, panoramic saliency considers the unique viewing behavior of HMD users, where attention is distributed across the equatorial region of equirectangular frames. 
Their approach leverages PanoSalNet, trained on a specialized dataset of head orientation logs, to generate saliency maps that closely match actual user fixations in dynamic scenes. 
When these maps are merged with historical head orientation data using Long Short-Term Memory (LSTM) networks~\cite{hochreiter1997long}, the resulting model shows marked improvements in prediction accuracy, particularly during rapid head movements prompted by novel visual stimuli.

Building on these insights, TRACK~\cite{rondon2022track} addresses shortcomings in previous fusion strategies for multi-modal data through a Structural-RNN-inspired architecture~\cite{jain2016structural}. 
By adaptively balancing the influences of user trajectories and saliency cues, TRACK achieves state-of-the-art performance across a range of content types, including both focus-driven and exploratory videos. Its modular design reduces overfitting and preserves robust predictions over extended time horizons.

Despite these advances, saliency-based and trajectory-focused approaches often overlook the underlying cognitive processes that guide user behavior. Factors such as risk assessment, exploratory impulses, and shifts in priorities have yet to be fully modeled in head-movement prediction. In addition, domain-specific data or heuristics can limit the generalizability of these methods to varied scenarios.

Recognizing this gap, \cite{curtis2022toward} emphasizes aligning agent movements with more believable cognitive processes. Their framework tightly couples physical motion to higher-level reasoning, thereby ensuring that an agent’s actions feel contextually meaningful and psychologically plausible. 

Building on these ideas, our proposed method leverages advanced reasoning capabilities derived from VLMs and LLMs. While saliency-based systems and multi-modal architectures focus on predicting where users might look, our approach aims to elucidate why they make these choices. This enables virtual agents to replicate not merely the spatial patterns of head rotations, but also the underlying motivations driving them—an essential step toward creating genuinely believable interactions in virtual environments.

\section{Research Motivation}
Research on realistic virtual agent motion has traditionally focused on macro-level behaviors, such as crowd simulations~\cite{panayiotou2022ccp, charalambous2023greil, ji2024text} or trajectory prediction~\cite{yue2022human, guo2022end, lin2025progressive, wong2022view, mangalam2021goals}. While these efforts have yielded valuable insights into group dynamics and movement patterns, relatively little attention has been given to micro-level behaviors—those subtle, individual actions such as nuanced gestures or context-specific decision-making processes. Consequently, important aspects of an agent’s realistic presence in VR environments remain underexplored.

Among the various micro behaviors, an agent’s head rotation stands out as a crucial factor in providing realistic virtual experiences. Implementing natural head rotations, however, poses significant challenges because it depends on a complex interplay of environmental awareness, cognitive evaluation, and decision-making. Rather than explicitly modeling these intricate factors, recent studies have often opted for data-driven approaches or have relied on saliency-based methods~\cite{rondon2022track, nguyen2018your}, using approximate measures of human attention to simulate head rotations. Although these techniques can capture certain visually salient cues, their limited treatment of deeper cognitive processes restricts their adaptability. As a result, not only do they struggle to generalize across diverse situations, but they also deviate noticeably from plausible human head movement.

To overcome these limitations, we first performed Experiment 1 to understand the underlying rationale behind human head-rotation decisions through empirical data collection. Drawing on these insights, we then devised a VLM/LLM–based framework to simulate this decision-making process, ensuring robust performance across arbitrary scenarios without pre-recorded data. 

\section{Experiment 1}
The primary goal of our framework is to replicate the way real humans move their heads. 
To achieve this, we must first acquire a detailed understanding of how people turn their heads when navigating a VR setting, as well as the underlying reasoning (e.g., searching for landmarks, monitoring threats, or exploring novel objects) that drives those movements.

In Experiment 1, we conduct a user study to collect empirical data on human head-motion trajectories alongside the self-reported or inferred rationale behind each head movement. This data serves as a critical foundation: it not only reveals real-world patterns of head orientation but also provides insight into the contextual factors that guide such behaviors.
By analyzing these findings, we can better design our agent’s head-rotation logic and validate whether our VLM/LLM-based approach effectively emulates human-like decision processes in diverse navigation scenarios.

\subsection{Apparatus and test settings}
The study involved 20 participants, comprising 11 males and 9 females. All participants had prior experience with VR. The $\mu$ and $\sigma$ of age were $24.27\pm2.60$.
The experiment was conducted using an Oculus Quest2 headset, paired with controllers, and operated on a computer equipped with an RTX 3090 graphics card and an AMD Ryzen 7 3800XT processor.

The virtual environments were designed to reflect common real-world scenarios (e.g., crosswalk, shopping mall, café, street, and bus). 
To collect diverse data, we introduced two distinct experimental conditions:
\begin{itemize}
\item Minimal-Distraction Condition (MDC): This setup contained few or no attention-diverting elements, simulating a typical environment with relatively low levels of visual interference.
\item Attention-Provoking Condition (APC): This setup contained strategically placed objects intended to capture attention or obstruct the view, such as dynamic signage, unexpected obstacles, or motion-triggered distractions, designed to capture the participant's attention. 
\end{itemize}

By examining the data obtained under two distinct conditions, we aimed to determine how additional distractors impact cognitive processes and decision-making.

\subsection{Method and Procedure}
Experiment 1 was conducted with two experimenters. Upon arrival, each participant completed a consent form and a demographics questionnaire. Participants then underwent a training session, lasting up to 15 minutes, to familiarize themselves with the virtual environment. This session included a 5-minute overview of the experiment, followed by up to 10 minutes of free practice. Note that the environment, a normal city scene, was not used in the main experiment.

\sloppy
To ensure participant well-being and validate the experimental conditions, we conducted the Simulator Sickness Questionnaire~\cite{kennedy1993simulator} (SSQ) before and after the experiment. No statistically significant difference was observed in SSQ scores, indicating minimal discomfort. Additionally, the Simulation Task Load Index~\cite{harris2020development} (SIM-TLX) results indicated a low level of cognitive workload during the scenarios.

Following the practice session, participants were presented with ten virtual scenarios derived from two experimental conditions (MDC and APC) and five environment types (crosswalk, shopping mall, café, street, and bus). In each scenario, participants were instructed to complete a scenario-specific goal: \textbf{Crosswalk-}cross safely to the other side; \textbf{Shopping Mall-}move from one end of the mall to the opposite side; \textbf{Café-}locate and sit at a table by the window; \textbf{Street-}walk safely to the far end of the street; and \textbf{Bus-}find and sit in an empty seat at the back. They were allowed up to 60 seconds to complete each scenario and could request a 180-second break between scenarios. 
To mitigate potential order effects, each participant experienced these ten scenarios in a randomized sequence. 

During each scenario, the participant’s visual field was captured as a continuous video recording, while head rotations and other sensor data were logged at runtime. Upon completion of all scenarios, participants reviewed their recorded videos and provided a self-reported rationale for each significant head turn.
Specifically, experimenters asked whether the head turn was to focus on a particular object or to scan the environment more broadly, and then requested participants to explain their underlying reasons. 
To facilitate reliable responses, participants were presented with example prompts such as ``it caught my eye,” ``I was curious,” ``it seemed dangerous,” ``I wanted to better understand some information,” ``I needed to confirm something,” or ``it felt odd or out of place.” 

\subsection{User Review Results}
\begin{figure}[!h]
  \centering
  \includegraphics[width=0.5\linewidth]{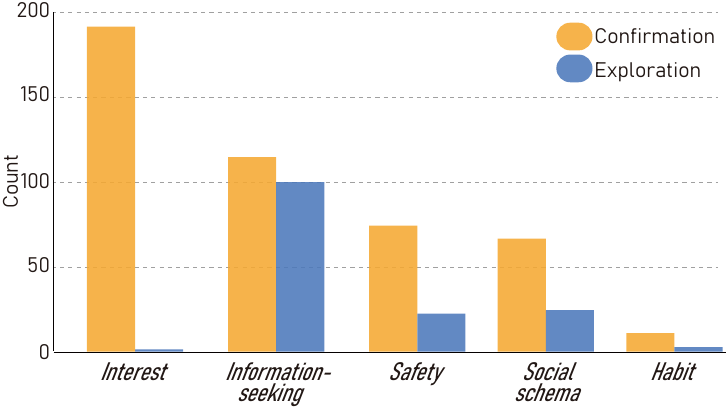}
  \caption{Categorized distribution of participants’ self-reported head-movement rationales.}
  \label{fig:MDC and APC}
\end{figure}

To uncover the underlying motivations for head movements, we categorized participants’ self-reported rationales into five as shown in~\autoref{fig:MDC and APC}. 
Three frequently cited motivations were \textit{interest}, \textit{information-seeking}, and \textit{safety}.
Interestingly, \textit{interest} tended to be referenced after participants had already oriented their heads, implying that they often noticed something peripherally before consciously deciding to focus on it.

Beyond these context-driven or curiosity-based explanations, participants also exhibited behaviors tied to \textit{habit} and \textit{social schema}. For instance, many individuals reflexively glanced toward their final destination as they walked, or automatically checked both directions at a crosswalk. Similarly, socially conditioned movements—such as scanning for a queue at a café counter or verifying the presence of other passengers on a bus—highlight the role of cultural and situational norms in influencing head rotations.

While these five categories provide insight into various cognitive and social triggers, our analysis revealed an additional benefit in grouping them based on the spatial extent of head movements. We use the term \textit{confirmation} to describe smaller, localized shifts within the existing field of view, commonly employed to verify details already noticed peripherally or to confirm the presence of known objects. By contrast, \textit{exploration} refers to broader, more pronounced rotations directed beyond the current line of sight, often associated with discovering new objects or scanning distant areas out of curiosity or concern. 
This distinction between confirmation and exploration helps differentiate between incremental checks triggered by prior awareness and more active, outward-directed searches for new information.

Taken together, these findings indicate that human head-rotation behaviors are driven not only by direct visual stimuli but also by learned behaviors, social norms, and situational awareness.

\begin{figure*}[!h]
  \centering
  \includegraphics[width=\textwidth]{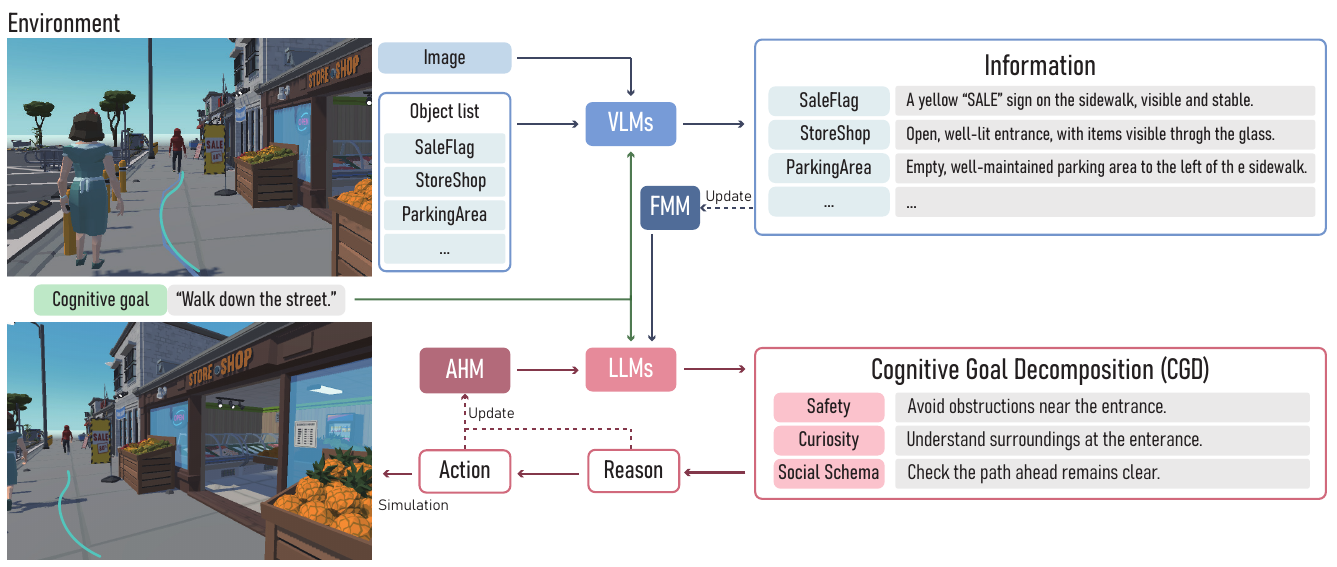}
  \caption{ViRAC Framework. This is an example of 'Street'. VLM identifies salient objects from the scene and updates the Foundational Memory Module (FMM). LLM then references both the Action History Module (AHM) and the FMM to decompose high-level cognitive goals into sub-tasks, guiding context-sensitive actions. }
  \label{fig:Architecture}
\end{figure*}

\section{Language-guided Framework}
As observed in E1, multiple interrelated factors influence the decision-making process behind head rotation. However, existing approaches often struggle to capture this complexity, instead relying on rigid heuristics or narrowly trained models~\cite{panayiotou2022ccp}. 

To address the limitation, we introduce \textbf{ViRAC}, a framework that combines a VLM and LLM to more plausibly emulate human cognition. By modeling the interplay among perception and decision-making modules, ViRAC can produce flexible, context-sensitive head movements in dynamic virtual environments.

\subsection{Perception Module}
The Perception Module processes the agent’s first-person view, detecting and describing objects to facilitate coherent and adaptive decision-making. It comprises two primary components: a VLM and a Foundational Memory Module (FMM).

Motivated by the need for robust, context-aware object recognition—without the overhead of manual annotation—the VLM automatically identifies salient objects in the agent’s field of view and generates coherent textual descriptions. Next, the system evaluates the relevance of these perceived objects by referencing the current cognitive goal and the memory state maintained by the FMM. Any objects deemed relevant are stored for extended recall, ensuring that the agent can seamlessly recognize them later and maintain continuity in complex tasks. 

This persistent memory mechanism is critical for decision-making in dynamic environments, where objects may temporarily leave the agent’s view or reappear in unexpected contexts. Detailed prompts used to guide the VLM are provided in the supplementary material.

\subsection{Decision-making Module}
The Decision-making Module bridges high-level objectives with adaptive, context-aware behaviors.
This comprises two core components: the Action History Module (AHM) and an LLM. 
Together, these components enable the agent to decompose tasks, track its past actions, and dynamically respond to changes in the environment.

The AHM records the actions executed by the agent, ensuring continuity across time and providing critical context for future decisions.  
Unlike traditional approaches that abstract actions into numerical parameters~\cite{cai2024bridging}, ViRAC preserves the semantic structure of actions in a human-like format. 
Specifically, the agent can \textit{``look at `object,'"} which entails focusing on a specific object in the field of view for detailed analysis or interaction, or \textit{``Search `direction,'"} which involves shifting attention toward a new direction to discover objects or areas beyond the current field of view. 
The AHM serves as a repository for all executed actions, providing a reference for the LLM during decision-making.

In our framework, the LLM decomposes a high-level cognitive goal into sub-tasks, ensuring a balance between exploration and task-focused objectives.  When constructing sub-goals, we exploit categorized rationales and behaviors obtained from Experiment 1 to form prompts for LLM decomposing. 
Furthermore, since LLMs often overemphasize safety or fail to generate nuanced and contextually appropriate sub-tasks without clear guidance, we exploit AHM and FMM to align subgoals with the given context. For example, in a shopping mall scenario, the LLM generates sub-goals such as ``scan nearby stores" or ``check escalator position."

\subsection{Framework Overview}
ViRAC operates in an iterative loop, allowing the agent to continuously \textit{perceive}, \textit{reason and action selection}, and \textit{update} its state as it navigates the virtual environment.

\paragraph{Perception} The VLM analyzes the agent’s current field of view (\(\mathbf{I}_t\)) and object list (\(\mathcal{O}_t\)), combined with the goal (\(G\)) and memory state (\(\mathcal{M}_t\)). VLM then produces a set of contextual object descriptions $\mathcal{D}_t = \mathcal{F}_{\text{VLM}}(\mathbf{I}_t, \mathcal{O}_t, G, \mathcal{M}_t)$. This process ensures that the agent maintains an up-to-date understanding of relevant objects.

\paragraph{Reasoning and Action Selection}
Given object descriptions (\(\mathcal{D}_t\)), cognitive goal (\(G\)), the agent's walking velocity (\(\mathbf{V}_t\)), and action history (\(\mathcal{H}_t\)), the LLM determines the most appropriate action $(\mathbf{a}_t) = \mathcal{F}_{\text{LLM}}(\mathcal{D}_t, G, \mathbf{V}_t, \mathcal{H}_t)$. This process ensures that ongoing objectives and prior behaviors influence the agent’s choices.

\paragraph{Environment Update}
Once the chosen action is executed, the environment ($\mathcal{E}$) adjusts the agent’s viewpoint ($\mathbf{I}_{t+1}$) and object list ($\mathcal{O}_{t+1}$). Specifically, $(\mathbf{I}_{t+1}, \mathcal{O}_{t+1}) = \mathcal{E}(\mathbf{a}_t).$

\paragraph{History Update}
The FMM records any new object descriptions, preserving essential information for future recall: $\mathcal{F}_{\text{FMM}}(\mathcal{D}_t)$. 
Simultaneously, the AMH records the executed action so that the system maintains continuity across tasks and time: $\mathcal{H}_{t+1} = \mathcal{H}_t \cup \{\mathbf{a}_t\}$.

\section{Evaluation}
To assess our framework’s ability to replicate user-like head movements in virtual environments, we compared our method against a baseline approach called \textit{Track}~\cite{rondon2022track} and user-generated motion data from Experiment 1 (henceforth ``\textit{Human}”). We used Dynamic Time Warping (DTW)~\cite{sakoe1978dynamic} as the primary metric for measuring how closely each simulated trajectory aligns with actual human behavior~\cite{cuturi2017soft, lerogeron2023approximating}.

\begin{table*}[ht]
  \centering
  \begin{tabular}{lcccccccccc}
    \toprule
    \multirow{2}{*}{Methods} & \multicolumn{2}{c}{Bus} & \multicolumn{2}{c}{Café} & \multicolumn{2}{c}{Crossing} & \multicolumn{2}{c}{Mall} & \multicolumn{2}{c}{Street}\\
    \cmidrule(lr){2-3}
    \cmidrule(lr){4-5}
    \cmidrule(lr){6-7}
    \cmidrule(lr){8-9}
    \cmidrule(lr){10-11}
            & MDC & APC & MDC & APC & MDC & APC & MDC & APC & MDC & APC \\
    \midrule
    Track & 0.3815 & \textbf{\underline{0.3888}} & 0.5840 & 0.5519 & 0.6510 & 0.7252 & 0.6645 & 0.6342 & 0.6165 & 0.6698 \\
    ViRAC (Ours) & \textbf{\underline{0.3082}} & 0.5861 & \textbf{\underline{0.5681}} & \textbf{\underline{0.4723}} & \textbf{\underline{0.5478}} & \textbf{\underline{0.6887}} & \textbf{\underline{0.4409}} & \textbf{\underline{0.4687}} & \textbf{\underline{0.4003}} & \textbf{\underline{0.3852}} \\
    \bottomrule
  \end{tabular}
  \caption{ Normalized DTW results for Track and ViRAC (Ours) across five scenario types under both Minimal-Distraction (MDC) and Attention-Provoking (APC) conditions. Lower scores indicate a closer match to the human head-rotation data. The lowest score in each column is bold and underlined.}
  \label{tab:Comparison}
\end{table*}

\subsection{Objective Evaluation --- Method}
We focused on five distinct virtual scenarios—Bus, Crossing, Café, Street, and Mall—under two conditions: MDC and APC, as used in Experiment 1. For each scenario–condition pair, we generated five runs of agent head-rotation trajectories using two different methods of \textit{ViRAC} and \textit{Track}. 
Human body trajectories collected in Experiment 1 were used as the agent's body position. 

Head rotations were encoded as quaternions because they efficiently capture 3D orientation. To quantify how well each method’s output matched \textit{Human} data, we computed the angular distance between any two quaternions \(q_1\) and \(q_2\) as
\begin{equation}
    d(q_1, q_2) = 2 \cdot \arccos\left(\left| \text{dot}(q_1, q_2) \right|\right),
\end{equation}
where \( \text{dot}(q_1, q_2) \) denotes the dot product of the normalized quaternions. 
We then employed DTW to optimally align the temporal sequences (human vs. model) to minimize the cumulative angular distance. Finally, we normalized these DTW scores by the average sequence length, allowing for fair comparisons even if the trajectories differed slightly in duration.

Identical prompts were used for both the MDC and APC conditions in ViRAC. This approach ensures that any differences in DTW scores arise from how each method handles the changing visual complexity rather than from diverging textual instructions.

\subsection{Objective Evaluation --- Results}

\autoref{tab:Comparison} provides the normalized DTW scores for both the \textit{Track} and our \textit{ViRAC} framework, with lower values denoting closer similarity to the \textit{Human} head-rotation

In most scenarios, \textit{ViRAC} achieved lower DTW scores than \textit{Track}, indicating trajectories that more closely resembled the \textit{Human} data. This suggests that combining first-person visual context with a language-driven cognitive model leads to more naturalistic head movements.

Environments like the \textit{Mall} and \textit{Crosswalk} posed considerable challenges due to frequent scene changes and the presence of multiple salient objects. In these dynamic settings, \textit{ViRAC} showed notably better performance, suggesting that the framework adapts well to visually complex or rapidly evolving contexts. 

Despite these gains, the \textit{Bus} scenario revealed an interesting limitation: \textit{Track} outperformed \textit{ViRAC} primarily due to a distinctive Santa Claus character that consistently drew participants’ attention. Because the LLM-based approach did not interpret Santa as a noteworthy element, \textit{ViRAC} failed to replicate the user behavior of focusing on this distractor. One plausible explanation is that the language model’s scene analysis overlooked the novelty or social relevance of the Santa figure, particularly if the model was not prompted to consider unusual or context-specific objects.

Although this shortcoming reduced the model’s overall performance in that scenario, it also reveals possible avenues for improvement. Beyond prompt engineering, a more iterative, context-aware scene analysis could allow the LLM to revisit its initial assessments, incorporate domain-specific knowledge, and dynamically assign greater importance to atypical objects. Such iterative refinement could improve the robustness of the \textit{ViRAC} in diverse scenarios.

\section{Subjective Evaluation --- Method}

\begin{figure}[t]
  \centering
  \includegraphics[width=0.5\linewidth]{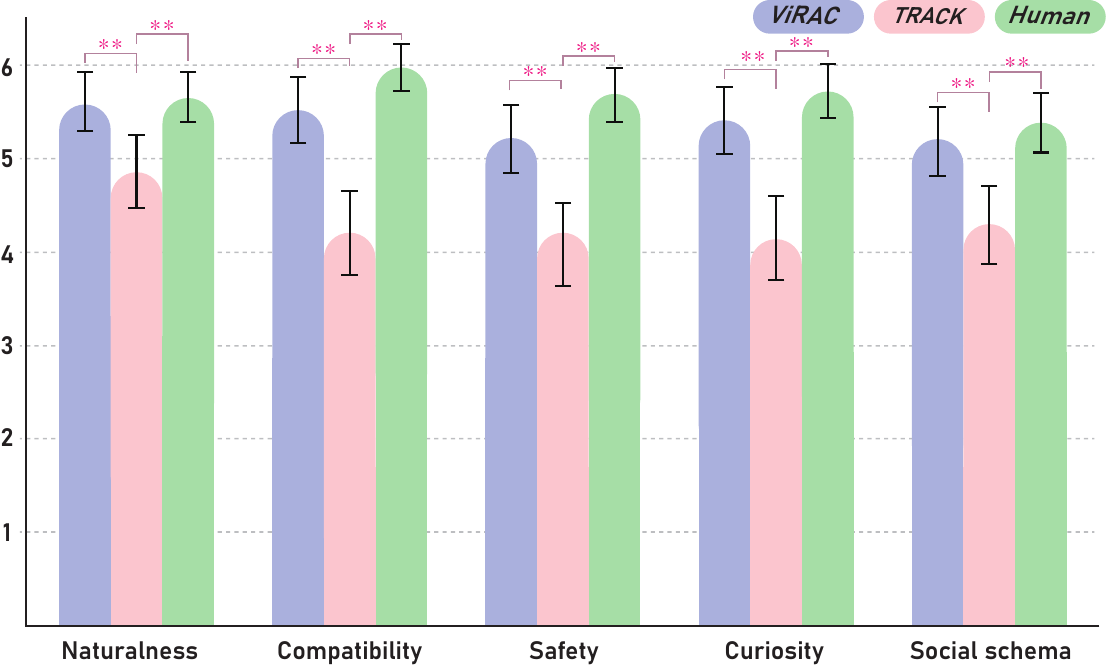}
  \caption{Mean scores and standard deviations for each metric, rated on a seven-point Likert scale. Higher values denote more favorable judgments. The brackets indicate statistically significant differences (**:$p$ < 0.01). \textit{ViRAC} achieves results comparable to \textit{Human} in most metrics and consistently outperforms \textit{Track}.}
  \label{fig:User study}
\end{figure}

To complement the objective analysis, we conducted a user study in which participants evaluated head-movement videos derived from Experiment 1 and the objective evaluation procedure.
For each scenario–condition pair (Bus, Crossing, Café, Street, Mall under MDC and APC), we identified three representative runs, omitting any trajectories that closely mirrored others. 
This yielded 30 distinct scenario–condition videos per method (\textit{ViRAC}, \textit{Track}, and \textit{Human}), for a total of 90 videos.
Each clip was shown in randomized order, accompanied by a concise text describing the agent’s actions and the rationale behind them. After watching, participants rated each video on five metrics using a seven-point Likert scale (0 = strongly negative, 6 = strongly positive):
\begin{itemize}
\item Naturalness: How natural do the agent’s head movements appear?
\item Compatibility: How well do the agent’s movements align with its stated goal?
\item Safety: How effectively does the agent detect potential dangers and navigate accordingly?
\item Curiosity: How well does the agent notice and respond to interesting elements in its environment?
\item Social Schema: How closely does the agent’s behavior follow social norms and conventions?
\end{itemize}

\section{Subjective Evaluation --- Results}
We used the Friedman test to detect statistical significance among three methods across five subjective metrics. 
This nonparametric test was chosen because our data did not satisfy the normality assumptions required for parametric alternatives. For post-hoc pairwise comparisons, we conducted Wilcoxon signed-rank tests.

As summarized in~ \autoref{fig:User study}, \textit{ViRAC} consistently achieved performance statistically comparable to \textit{Human} and outperformed \textit{Track} in all cases. 
For Naturalness ($\chi^2 = 9.80, p <0.01$), \textit{ViRAC} scored similarly to \textit{Human}, while significantly surpassing \textit{Track} ($Z=-11.32, p<0.01$). 
For Compatibility ($\chi^2 = 14.92, p <0.01$), \textit{ViRAC} again scored similarly to \textit{Human}, while significantly surpassing \textit{Track} ($Z=-13.54, p<0.01$).
For Safety ($\chi^2 = 13.40, p <0.01$), \textit{ViRAC} scored similarly to \textit{Human} while significantly surpassing \textit{Track} ($Z=-13.48, p<0.01$).
For Curiosity ($\chi^2 = 10.40, p <0.01$), \textit{ViRAC} scored similarly to \textit{Human} while significantly surpassing \textit{Track} ($Z=-11.49, p<0.01$).
For Social Schema ($\chi^2 = 12.20, p <0.01$), \textit{ViRAC} scored similarly to \textit{Human} while significantly surpassing  \textit{Track} ($Z=-13.00, p<0.01$).

These findings suggest that \textit{ViRAC} performs at a level close to actual human head-rotation behavior, specifically in terms of naturalness, compatibility with the given task, safety-focused detection, curiosity-driven engagement, and socially normative responses. 



\section{Conclusion}
\begin{figure*}[t]
  \centering
  \includegraphics[width=1\linewidth]{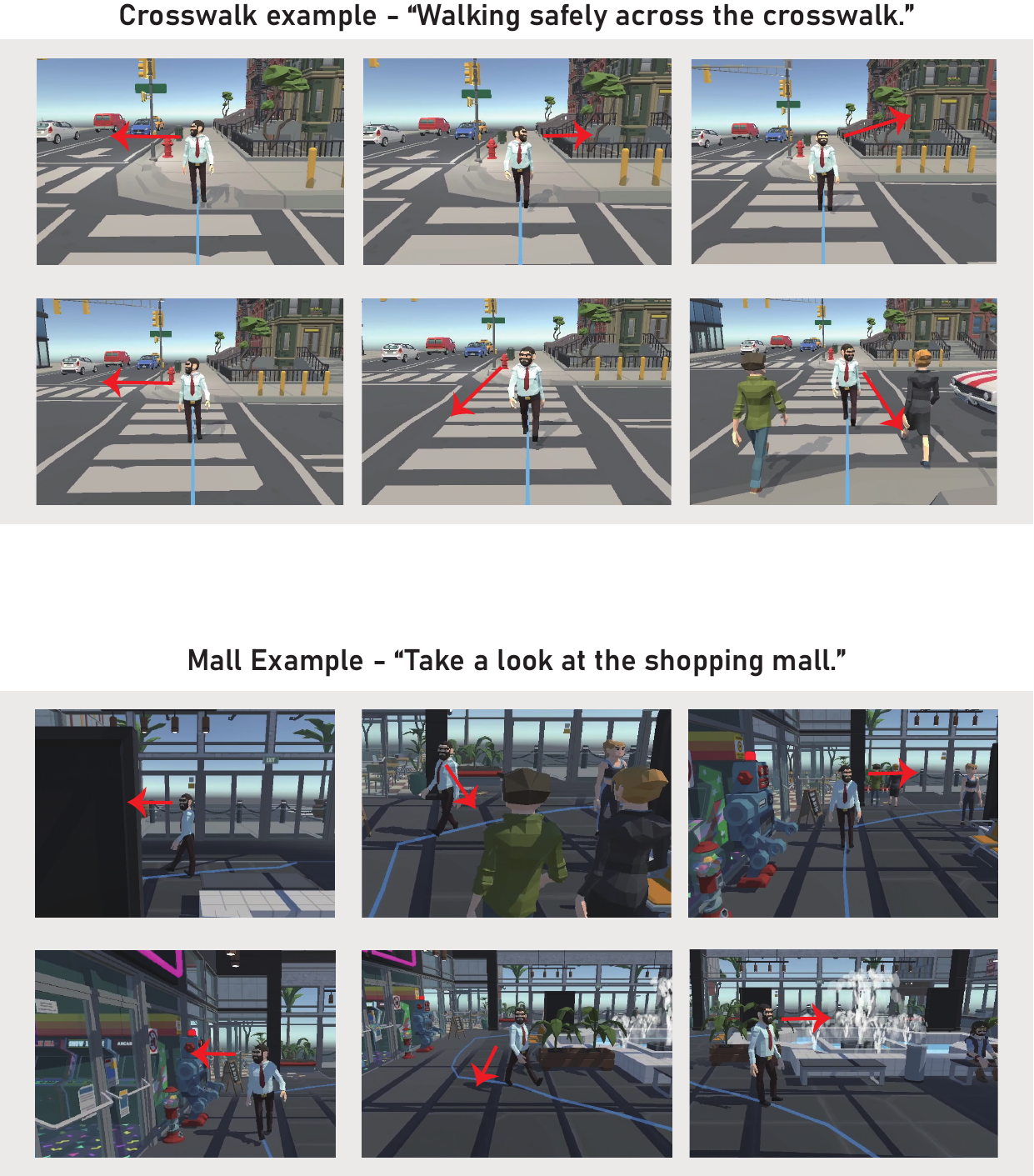}
  \caption{Sample frames from the crosswalk and mall scenarios.}
  \label{fig:Example1}
\end{figure*}

\begin{figure*}[t]
  \centering
  \includegraphics[width=1\linewidth]{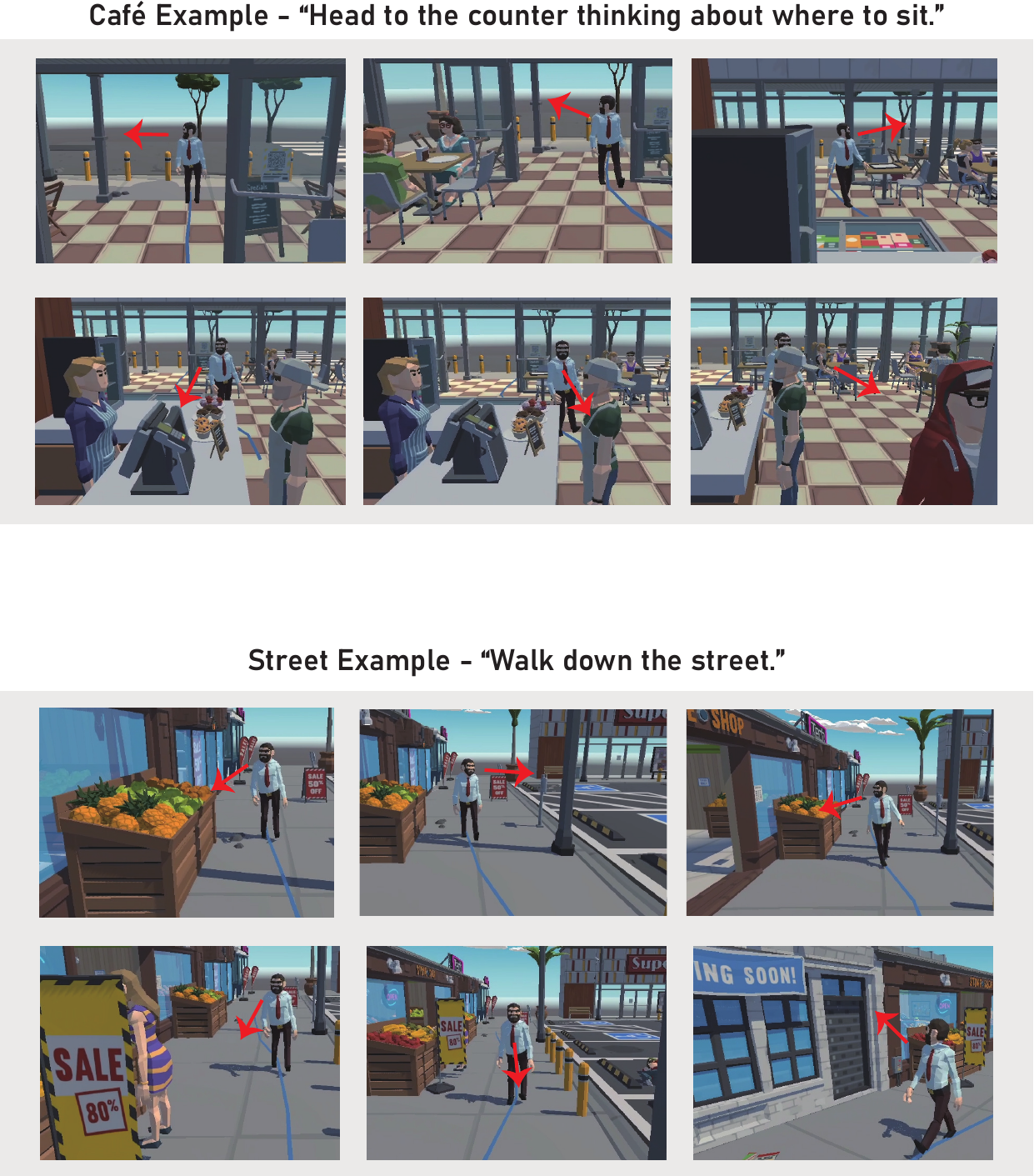}
  \caption{Sample frames from the café and street scenarios.}
  \label{fig:Example2}
\end{figure*}

We have introduced ViRAC, a language-guided framework for generating human-like head movements in virtual agents.
By unifying VLM and LLM, ViRAC interprets environmental cues with an unprecedented depth of reasoning, enabling more convincing and context-sensitive agent behaviors than earlier, purely data-driven or saliency-based methods. 
Our experiments demonstrate that ViRAC improves upon the TRACK method in aligning agent head rotations with human ground-truth data, thereby advancing the realism of virtual environments. Sample frames showing agent's head rotation generated by ViRAC are presented in~\autoref{fig:Example1} and ~\autoref{fig:Example2}.

Despite these advances, several limitations invite future exploration. First, ViRAC currently relies on visual data alone, limiting its adaptability in scenarios where non-visual cues or multimodal inputs—such as audio or haptic feedback—play a significant role. Integrating additional sensory streams could broaden the framework’s applicability and further enhance realism.
Second, ViRAC focuses on head-movement determination while omitting path planning, which remains crucial for tasks requiring coherent locomotion or close proximity object interactions. Coupling ViRAC’s perceptual and cognitive modules with a robust navigation system may help unify head rotation with locomotion, producing fully coordinated agent actions. Lastly, while large-scale language models offer rich contextual knowledge, their biases, and incomplete domain coverage can yield occasional oversights (e.g., ignoring atypical distractors). Refining prompt engineering, expanding training sets, or incorporating scene-adaptive updates may help address these gaps.

\bibliographystyle{unsrt}  
\bibliography{references}  

\begin{thebibliography}{10}

\bibitem{lerner2010context}
Alon Lerner, Yiorgos Chrysanthou, Ariel Shamir, and Daniel Cohen-Or.
\newblock Context-dependent crowd evaluation.
\newblock In {\em Computer Graphics Forum}, volume~29, pages 2197--2206. Wiley Online Library, 2010.

\bibitem{steel2010context}
Travis Steel, Dane Kuiper, and RZ~Wenkstern.
\newblock Context-aware virtual agents in open environments.
\newblock In {\em 2010 Sixth International Conference on Autonomic and Autonomous Systems}, pages 90--96. IEEE, 2010.

\bibitem{narang2016pedvr}
Sahil Narang, Andrew Best, Tanmay Randhavane, Ari Shapiro, and Dinesh Manocha.
\newblock Pedvr: Simulating gaze-based interactions between a real user and virtual crowds.
\newblock In {\em Proceedings of the 22nd ACM conference on virtual reality software and technology}, pages 91--100, 2016.

\bibitem{curtis2022toward}
Cassidy Curtis, Sigurdur~Orn Adalgeirsson, Horia~Stefan Ciurdar, Peter McDermott, JD~Vel{\'a}squez, W~Bradley Knox, Alonso Martinez, Dei Gaztelumendi, Norberto~Adrian Goussies, Tianyu Liu, et~al.
\newblock Toward believable acting for autonomous animated characters.
\newblock In {\em Proceedings of the 15th ACM SIGGRAPH Conference on Motion, Interaction and Games}, pages 1--15, 2022.

\bibitem{xu2018gaze}
Yanyu Xu, Yanbing Dong, Junru Wu, Zhengzhong Sun, Zhiru Shi, Jingyi Yu, and Shenghua Gao.
\newblock Gaze prediction in dynamic 360 immersive videos.
\newblock In {\em proceedings of the IEEE Conference on Computer Vision and Pattern Recognition}, pages 5333--5342, 2018.

\bibitem{zhu2020learning}
Yucheng Zhu, Guangtao Zhai, Xiongkuo Min, and Jiantao Zhou.
\newblock Learning a deep agent to predict head movement in 360-degree images.
\newblock {\em ACM Transactions on Multimedia Computing, Communications, and Applications (TOMM)}, 16(4):1--23, 2020.

\bibitem{yang2021hierarchical}
Li~Yang, Mai Xu, Yichen Guo, Xin Deng, Fangyuan Gao, and Zhenyu Guan.
\newblock Hierarchical bayesian lstm for head trajectory prediction on omnidirectional images.
\newblock {\em IEEE Transactions on Pattern Analysis and Machine Intelligence}, 44(11):7563--7580, 2021.

\bibitem{rondon2022track}
MFR Rondon, L~Sassatelli, R~Aparicio-Pardo, and F~Precioso.
\newblock Track: A new method from a re-examination of deep architectures for head motion prediction in 360° videos.
\newblock {\em IEEE Transactions on Pattern Analysis and Machine Intelligence}, 44(9):5681--5699, 2022.

\bibitem{hsu2020generalized}
Yen-Chang Hsu, Yilin Shen, Hongxia Jin, and Zsolt Kira.
\newblock Generalized odin: Detecting out-of-distribution image without learning from out-of-distribution data.
\newblock In {\em Proceedings of the IEEE/CVF conference on computer vision and pattern recognition}, pages 10951--10960, 2020.

\bibitem{yang2024qwen2}
An~Yang, Baosong Yang, Beichen Zhang, Binyuan Hui, Bo~Zheng, Bowen Yu, Chengyuan Li, Dayiheng Liu, Fei Huang, Haoran Wei, et~al.
\newblock Qwen2. 5 technical report.
\newblock {\em arXiv preprint arXiv:2412.15115}, 2024.

\bibitem{havrilla2024glore}
Alex Havrilla, Sharath Raparthy, Christoforus Nalmpantis, Jane Dwivedi-Yu, Maksym Zhuravinskyi, Eric Hambro, and Roberta Raileanu.
\newblock Glore: When, where, and how to improve llm reasoning via global and local refinements.
\newblock {\em arXiv preprint arXiv:2402.10963}, 2024.

\bibitem{qian2016optimizing}
Feng Qian, Lusheng Ji, Bo~Han, and Vijay Gopalakrishnan.
\newblock Optimizing 360 video delivery over cellular networks.
\newblock In {\em Proceedings of the 5th Workshop on All Things Cellular: Operations, Applications and Challenges}, pages 1--6, 2016.

\bibitem{duanmu2017prioritized}
Fanyi Duanmu, Eymen Kurdoglu, S~Amir Hosseini, Yong Liu, and Yao Wang.
\newblock Prioritized buffer control in two-tier 360 video streaming.
\newblock In {\em Proceedings of the Workshop on Virtual Reality and Augmented Reality Network}, pages 13--18, 2017.

\bibitem{nguyen2018your}
Anh Nguyen, Zhisheng Yan, and Klara Nahrstedt.
\newblock Your attention is unique: Detecting 360-degree video saliency in head-mounted display for head movement prediction.
\newblock In {\em Proceedings of the 26th ACM international conference on Multimedia}, pages 1190--1198, 2018.

\bibitem{hochreiter1997long}
S~Hochreiter.
\newblock Long short-term memory.
\newblock {\em Neural Computation MIT-Press}, 1997.

\bibitem{jain2016structural}
Ashesh Jain, Amir~R Zamir, Silvio Savarese, and Ashutosh Saxena.
\newblock Structural-rnn: Deep learning on spatio-temporal graphs.
\newblock In {\em Proceedings of the ieee conference on computer vision and pattern recognition}, pages 5308--5317, 2016.

\bibitem{panayiotou2022ccp}
Andreas Panayiotou, Theodoros Kyriakou, Marilena Lemonari, Yiorgos Chrysanthou, and Panayiotis Charalambous.
\newblock Ccp: Configurable crowd profiles.
\newblock In {\em ACM SIGGRAPH 2022 conference proceedings}, pages 1--10, 2022.

\bibitem{charalambous2023greil}
Panayiotis Charalambous, Julien Pettre, Vassilis Vassiliades, Yiorgos Chrysanthou, and Nuria Pelechano.
\newblock Greil-crowds: crowd simulation with deep reinforcement learning and examples.
\newblock {\em ACM Transactions on Graphics (TOG)}, 42(4):1--15, 2023.

\bibitem{ji2024text}
Xuebo Ji, Zherong Pan, Xifeng Gao, and Jia Pan.
\newblock Text-guided synthesis of crowd animation.
\newblock In {\em ACM SIGGRAPH 2024 Conference Papers}, pages 1--11, 2024.

\bibitem{yue2022human}
Jiangbei Yue, Dinesh Manocha, and He~Wang.
\newblock Human trajectory prediction via neural social physics.
\newblock In {\em European conference on computer vision}, pages 376--394. Springer, 2022.

\bibitem{guo2022end}
Ke~Guo, Wenxi Liu, and Jia Pan.
\newblock End-to-end trajectory distribution prediction based on occupancy grid maps.
\newblock In {\em Proceedings of the IEEE/CVF Conference on Computer Vision and Pattern Recognition}, pages 2242--2251, 2022.

\bibitem{lin2025progressive}
Xiaotong Lin, Tianming Liang, Jianhuang Lai, and Jian-Fang Hu.
\newblock Progressive pretext task learning for human trajectory prediction.
\newblock In {\em European Conference on Computer Vision}, pages 197--214. Springer, 2025.

\bibitem{wong2022view}
Conghao Wong, Beihao Xia, Ziming Hong, Qinmu Peng, Wei Yuan, Qiong Cao, Yibo Yang, and Xinge You.
\newblock View vertically: A hierarchical network for trajectory prediction via fourier spectrums.
\newblock In {\em European Conference on Computer Vision}, pages 682--700. Springer, 2022.

\bibitem{mangalam2021goals}
Karttikeya Mangalam, Yang An, Harshayu Girase, and Jitendra Malik.
\newblock From goals, waypoints \& paths to long term human trajectory forecasting.
\newblock In {\em Proceedings of the IEEE/CVF International Conference on Computer Vision}, pages 15233--15242, 2021.

\bibitem{kennedy1993simulator}
Robert~S Kennedy, Norman~E Lane, Kevin~S Berbaum, and Michael~G Lilienthal.
\newblock Simulator sickness questionnaire: An enhanced method for quantifying simulator sickness.
\newblock {\em The international journal of aviation psychology}, 3(3):203--220, 1993.

\bibitem{harris2020development}
David Harris, Mark Wilson, and Samuel Vine.
\newblock Development and validation of a simulation workload measure: the simulation task load index (sim-tlx).
\newblock {\em Virtual Reality}, 24(4):557--566, 2020.

\bibitem{cai2024bridging}
Wenzhe Cai, Siyuan Huang, Guangran Cheng, Yuxing Long, Peng Gao, Changyin Sun, and Hao Dong.
\newblock Bridging zero-shot object navigation and foundation models through pixel-guided navigation skill.
\newblock In {\em 2024 IEEE International Conference on Robotics and Automation (ICRA)}, pages 5228--5234. IEEE, 2024.

\bibitem{sakoe1978dynamic}
Hiroaki Sakoe and Seibi Chiba.
\newblock Dynamic programming algorithm optimization for spoken word recognition.
\newblock {\em IEEE transactions on acoustics, speech, and signal processing}, 26(1):43--49, 1978.

\bibitem{cuturi2017soft}
Marco Cuturi and Mathieu Blondel.
\newblock Soft-dtw: a differentiable loss function for time-series.
\newblock In {\em International conference on machine learning}, pages 894--903. PMLR, 2017.

\bibitem{lerogeron2023approximating}
Hugo Lerogeron, Romain Picot-Clemente, Alain Rakotomamonjy, and Laurent Heutte.
\newblock Approximating dtw with a convolutional neural network on eeg data.
\newblock {\em arXiv preprint arXiv:2301.12873}, 2023.

\end{thebibliography}


\begin{thebibliography}{1}

\bibitem{kour2014real}
George Kour and Raid Saabne.
\newblock Real-time segmentation of on-line handwritten arabic script.
\newblock In {\em Frontiers in Handwriting Recognition (ICFHR), 2014 14th
  International Conference on}, pages 417--422. IEEE, 2014.

\bibitem{kour2014fast}
George Kour and Raid Saabne.
\newblock Fast classification of handwritten on-line arabic characters.
\newblock In {\em Soft Computing and Pattern Recognition (SoCPaR), 2014 6th
  International Conference of}, pages 312--318. IEEE, 2014.

\bibitem{hadash2018estimate}
Guy Hadash, Einat Kermany, Boaz Carmeli, Ofer Lavi, George Kour, and Alon
  Jacovi.
\newblock Estimate and replace: A novel approach to integrating deep neural
  networks with existing applications.
\newblock {\em arXiv preprint arXiv:1804.09028}, 2018.

\end{thebibliography}


\renewcommand{\thesection}{\Alph{section}} 
\renewcommand{\thesubsection}{\thesection.\arabic{subsection}} 

\begin{figure*}
\centering
\includegraphics[width=\textwidth,height=\textheight-10,keepaspectratio]{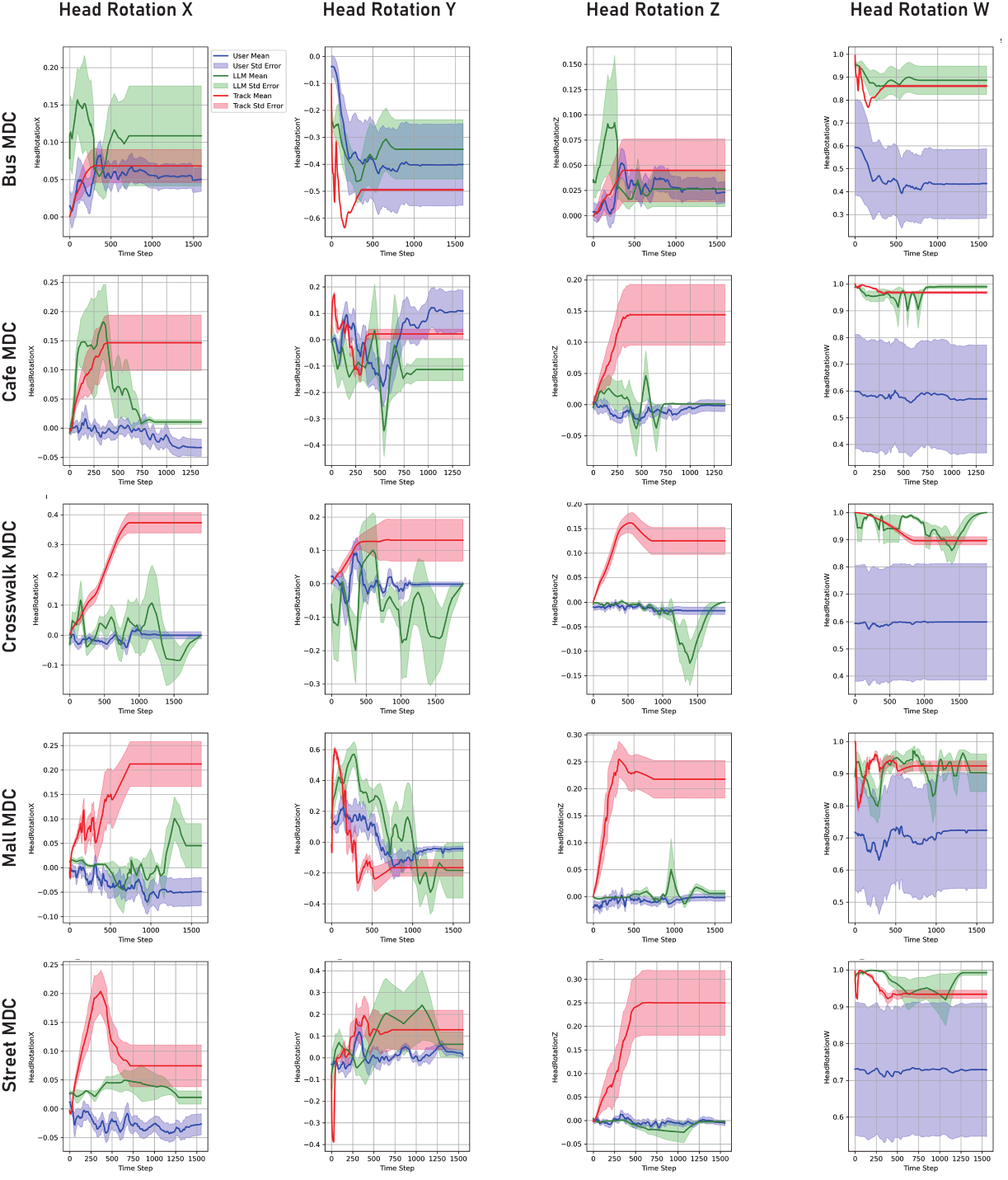}
\caption{Head-rotation data (expressed as quaternions) under MDC across five different environments (Bus, café, crosswalk, Mall, and Street).}
\end{figure*}

\begin{figure*}
\centering
\includegraphics[width=\textwidth,height=\textheight,keepaspectratio]{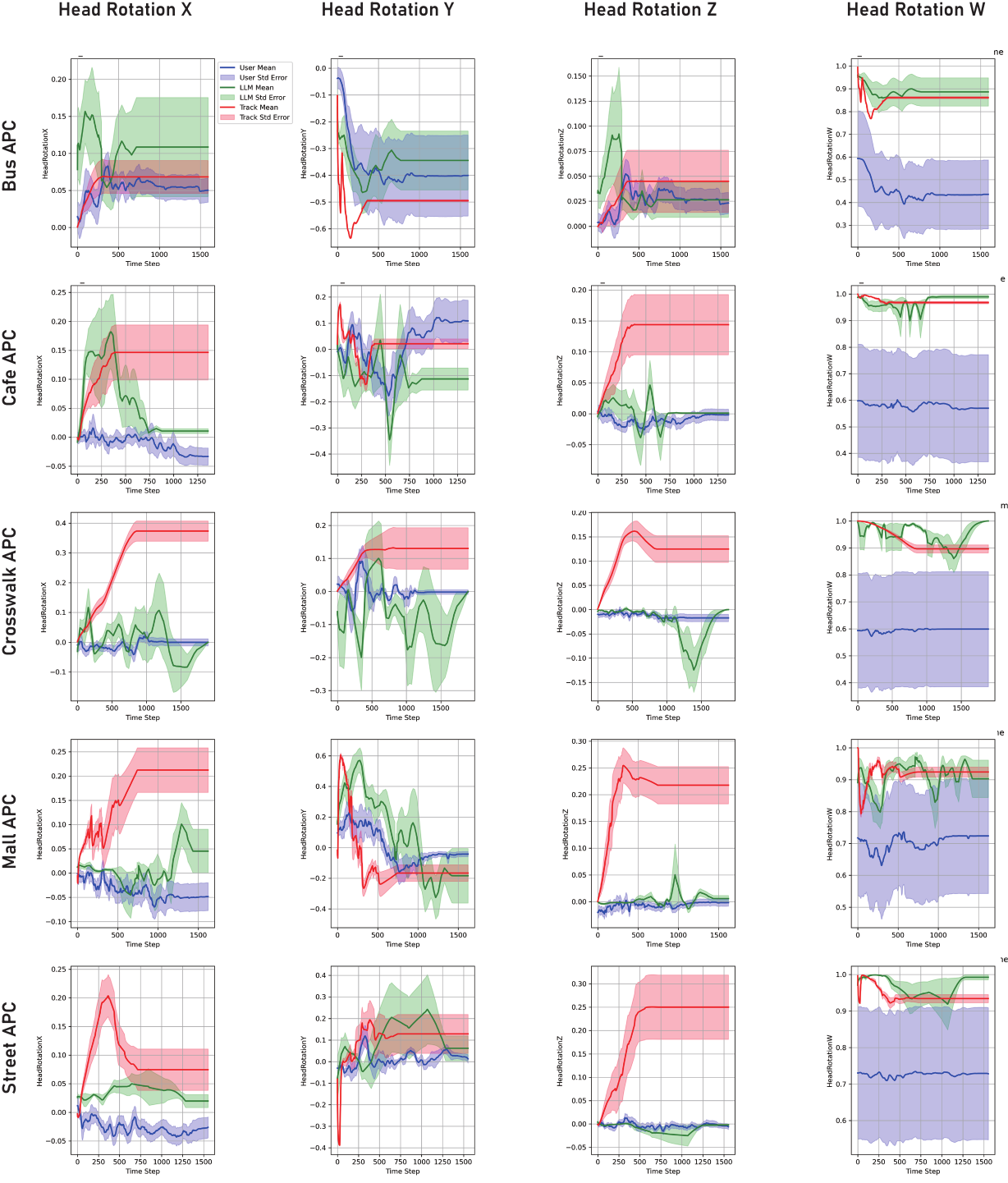}
\caption{Head-rotation data (expressed as quaternions) under APC across five different environments (Bus, café, crosswalk, Mall, and Street).}
\end{figure*}

\begin{figure*}
\centering
\includegraphics[width=\textwidth,height=\textheight,keepaspectratio]{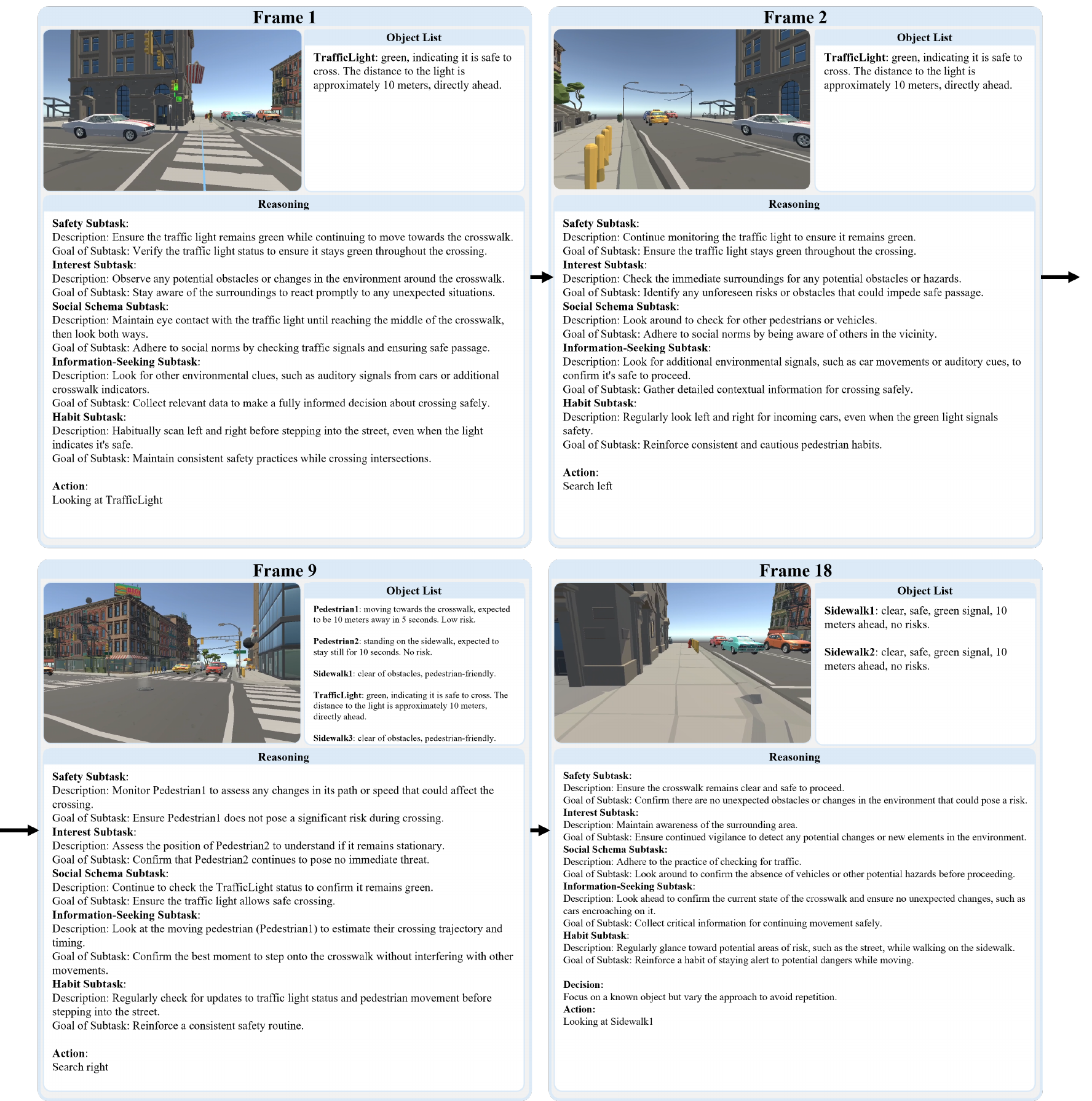}
\caption{The agent’s first-person view, a corresponding object list, and the LLM’s reasoning in crosswalk scenario.}
\end{figure*}

\begin{figure*}
\centering
\includegraphics[width=\textwidth,height=\textheight,keepaspectratio]{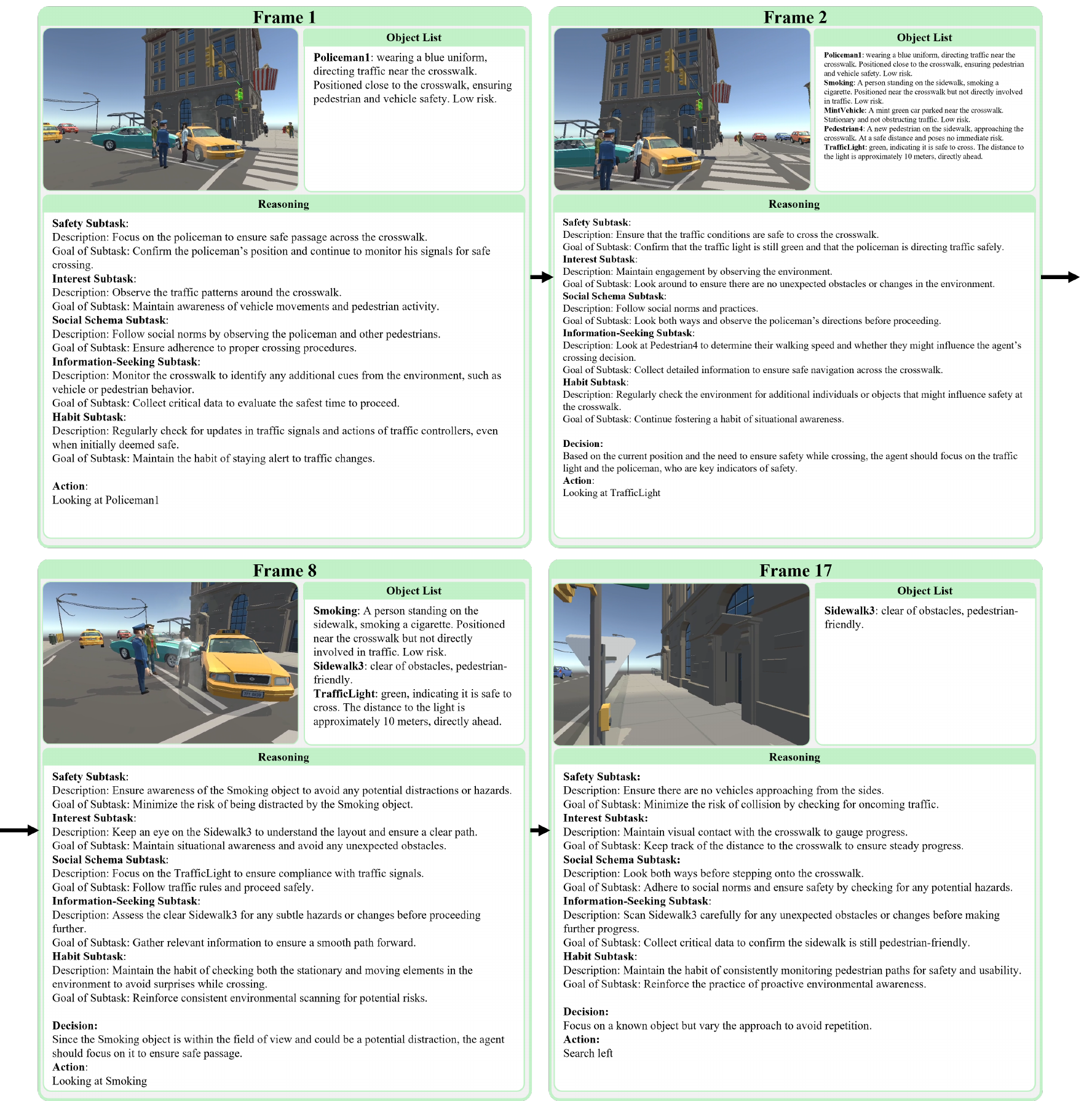}
\caption{The agent’s first-person view, a corresponding object list, and the LLM’s reasoning in crosswalk scenario.}
\end{figure*}

\begin{figure*}
\centering
\includegraphics[width=\textwidth,height=\textheight,keepaspectratio]{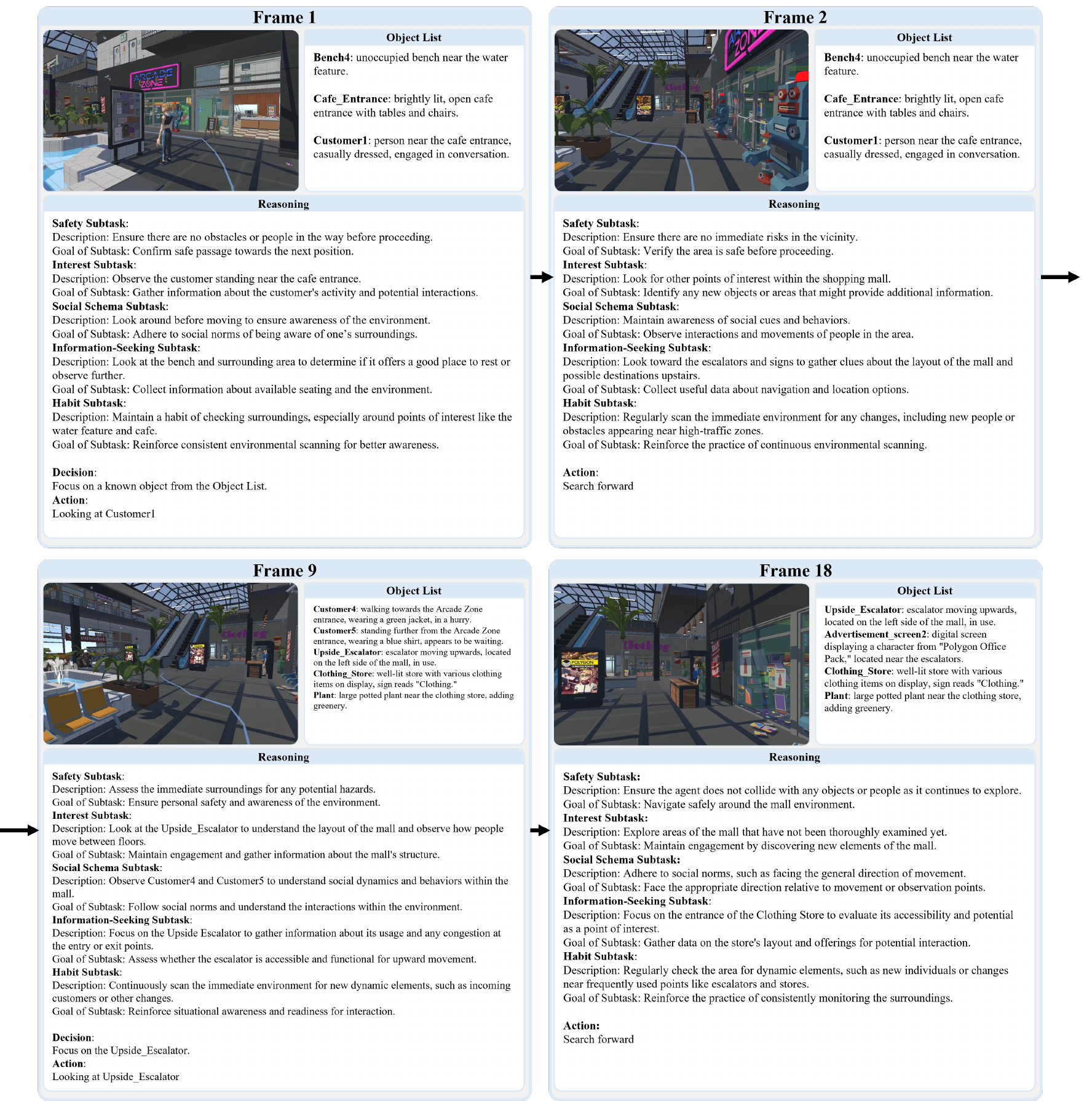}
\caption{The agent’s first-person view, a corresponding object list, and the LLM’s reasoning in mall scenario.}
\end{figure*}

\begin{figure*}
\centering
\includegraphics[width=\textwidth,height=\textheight,keepaspectratio]{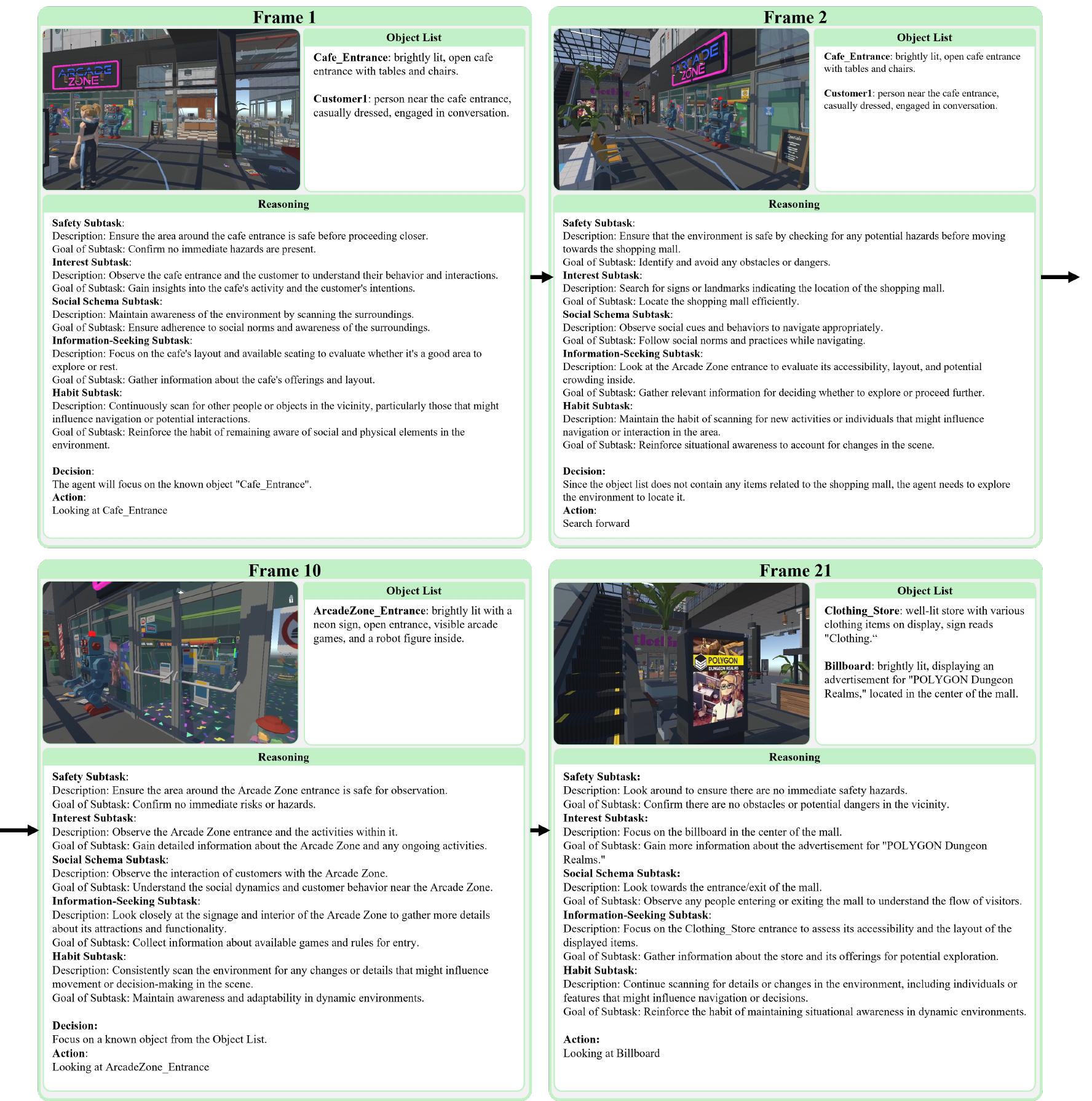}
\caption{The agent’s first-person view, a corresponding object list, and the LLM’s reasoning in mall scenario..}
\end{figure*}

\begin{figure*}
\centering
\includegraphics[width=\textwidth,height=\textheight,keepaspectratio]{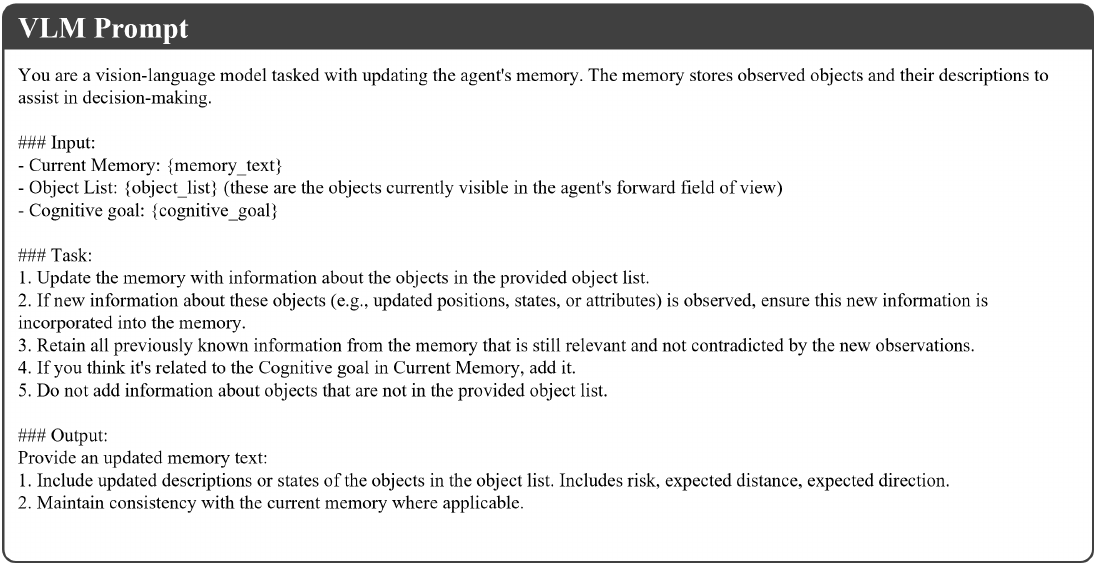}
\caption{VLM prompt detailing how to update the agent’s memory with newly observed objects and their attributes. }
\end{figure*}

\begin{figure*}
\centering
\includegraphics[width=\textwidth,height=\textheight,keepaspectratio]{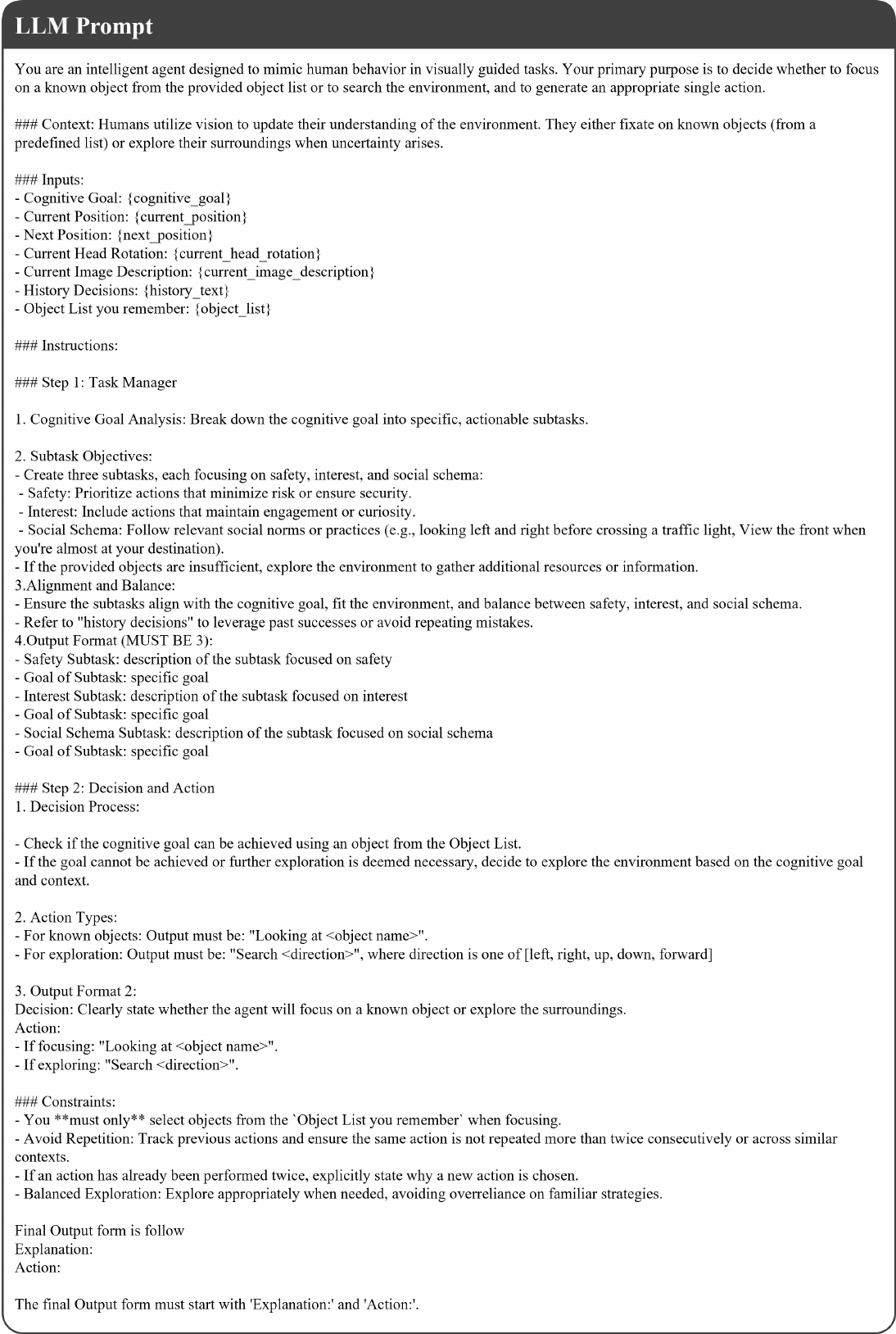}
\caption{LLM prompt detailing how to decompose the goal and select the action. }
\end{figure*}

\end{document}